\documentclass[prd,preprintnumbers,floatfix,aps,notitlepage,nofootinbib,twocolumn,showpacs,amssymb]{revtex4-2}
\usepackage{amsmath,amsfonts,bm}
\usepackage{graphicx}
\usepackage{cancel}
\usepackage{booktabs} 
\usepackage{array}    
\usepackage[colorlinks, linkcolor={red},citecolor={blue}]{hyperref}
\usepackage{xcolor}

\begin{document}
\title{On Schwarzschild black hole singularity formation}
\author{Jorge Ovalle}
\email[]{jorge.ovalle@physics.slu.cz}
\affiliation{Research Centre for Theoretical Physics and Astrophysics,
	Institute of Physics, Silesian University in Opava, CZ-746 01 Opava,
	Czech Republic.}	
\author{Roberto Casadio}
\email{casadio@bo.infn.it}
\affiliation{Dipartimento di Fisica e Astronomia,
	Alma Mater Universit\`a di Bologna,
	40126 Bologna, Italy
	\\
	Istituto Nazionale di Fisica Nucleare, I.S.~FLaG
	Sezione di Bologna, 40127 Bologna, Italy
	\\
	Alma Mater Research Center on Applied Mathematics (AM$^2$),
	Via Saragozza 8, 40123 Bologna, Italy}
\author{Alexander Kamenshchik}
\email{kamenshchik@bo.infn.it}
\affiliation{Dipartimento di Fisica e Astronomia,
	Alma Mater Universit\`a di Bologna,
	40126 Bologna, Italy
	\\
	Istituto Nazionale di Fisica Nucleare, I.S.~FLaG
	Sezione di Bologna, 40127 Bologna, Italy}

\begin{abstract}
\noindent We examine whether the Schwarzschild black hole can emerge as the continuous end state of gravitational collapse from a non-singular configuration. Employing a time dependent extension of the regular Schwarzschild metric, we track the evolution of the geometry during collapse and find that the process cannot remain continuous. The metric function develops a discontinuity at the origin, marking a breakdown of spacetime smoothness, an effect identified as “Minkowski breaking.” Before the Schwarzschild point source can form at $r=0$, curvature singularities appear and the Cauchy horizon disappears. These results strongly suggest that spacetime may not evolve smoothly toward the Schwarzschild geometry. Instead, the formation of a Schwarzschild black hole appears to entail a discrete change in the structure of spacetime, pointing to the need for a noncontinuous, possibly quantized, framework to describe the emergence or regularization of gravitational singularities.

		\end{abstract} 
\maketitle
%
%
%
\section{Introduction}

\noindent  In general relativity, the celebrated Schwarzschild solution arises as the answer to a problem defined by two conditions: (i) the existence of a vacuum, that is, the absence of any gravitational source, and (ii) the assumption of spherical symmetry. At first glance, these two requirements appear mutually contradictory, since a spherically symmetric spacetime presupposes the presence of a source that deforms Minkowski space in a specific manner, namely through the generators of spherical symmetry represented by three spacelike Killing vectors. Thus, imposing spherical symmetry (condition ii) implicitly demands the existence of a source, thereby contradicting condition (i). The standard resolution of this apparent paradox is to introduce a point-like mass ${\cal M}$ located precisely at the center of symmetry, $r=0$, which makes condition (i) strictly valid only for $r>0$. In summary, when the two conditions above are imposed, what we are truly doing is specifying a point-like mass as the gravitational source and then inquiring about the field it generates. It should come as no surprise, therefore, that this gravitational field diverges precisely at $r=0$. From a mathematical standpoint, however, this interpretation is only heuristic, since it has been shown~\cite{Geroch:1987aa} that the Einstein tensor cannot consistently accommodate a point-mass energy-momentum tensor~\cite{Balasin:1993fn}. In essence, the two conditions together describe the final field generated by a singularity resulting from total collapse, while entirely neglecting the dynamical process that leads to it. Consequently, any study focused on the dynamical process of collapse and its resulting singularity, rather than the static point source end-state, must abandon condition (i). This is a foundational prerequisite, for instance, for formulating a non-singular black hole (BH) model.

Numerous non-singular BH solutions have been proposed~\cite{Bardeen:1968qtr,Dymnikova:1992ux,Hayward:2005gi,BenAchour:2020gon,Bonanno:2023rzk}, yet the mechanisms by which such objects could emerge as the end-state of gravitational collapse remain poorly understood. The Penrose singularity theorem~\cite{Penrose:1964wq,Hawking:1970zqf,Hawking:1973uf} clarifies how these solutions can, in principle, exist by requiring the violation of at least one of the theorem’s assumptions, but it offers no insight into the dynamical processes that would actually produce them. A step toward this understanding was recently provided in Ref.~\cite{Ovalle:2025pue}, which analyzes gravitational collapse within the Schwarzschild BH. That work shows, without relying on any specific gravitational theory and without invoking perturbative methods~\cite{Poisson:1989zz,Ori:1991zz,Carballo-Rubio:2018pmi,Carballo-Rubio:2021bpr}, that a regular Schwarzschild BH will inevitably develop singularities, in full agreement with the strong cosmic censorship conjecture~\cite{Penrose:1969pc,Dafermos:2003wr,Dafermos:2017dbw,Hollands:2019whz}. Moreover, it demonstrates that such a non-singular BH configuration, of differentiability class ${\cal C}^N$ with $N>1$, can arise as a transient state only if the system avoids singularities inherent to the collapse kinematics, singularities that are absent in purely static configurations. This result underscores an important point: if a (transient) non-singular BH with a ${\cal C}^{N>1}$ geometry does form, it must be the outcome of an exceptionally constrained collapse, at least in the spherically symmetric case. Whether analogous dynamical singularities occur in axisymmetric collapse remains unknown, though there is little reason to expect a fundamentally different scenario.

Building upon these results, and given that we now possess a regular Schwarzschild configuration with an exceptionally rich internal structure, we are in a position to address a fundamental question: Can the Schwarzschild solution itself be formed through a continuous, dynamical process? In other words, rather than assuming a point source located at $r=0$ as the final outcome of collapse, we ask whether the evolution toward such a configuration can be described in a self-consistent manner. Under fairly general assumptions, we provide compelling evidence suggesting that the Schwarzschild solution may not emerge as the continuous end-state of gravitational collapse. More strikingly, we find that curvature singularities inevitably develop before the point source even begins to form at $r=0$, indicating that the spacetime continuum itself breaks down precisely at the very location where the Schwarzschild source would otherwise be expected to appear.

 \section{Regular Schwarzschild BHs.}
 \label{sec1}
 \noindent  To make this discussion self-contained, we first clarify what we mean by a revisited Schwarzschild BH~\cite{Ovalle:2024wtv}. Its geometry is described by the line element
 \begin{equation}
 	\label{metric}
 	ds^{2}
 	=-f(r)\,dt^{2}
 	+\frac{dr^{2}}{f(r)}
 	+r^{2} d\Omega^{2}, 
 \end{equation}
 with
\begin{equation}
	\label{mtransform}
	f(r)=\left\{
	\begin{array}{l}
		1-\frac{2\,m(r)}{r}
		\ ,
		\quad
		{\rm for}\
		0< r \leq h
		\\
		\\
		1-\frac{2\,{\cal M}}{r}
		\ ,
		\quad
		{\rm for}\
		r>h
		\ .
	\end{array}
	\right.
\end{equation}
 Here $m(r)$ is the Misner–Sharp mass, and
 \begin{equation}
 	\label{cond1}
 	{\cal M}\equiv m(h)=\frac{h}{2}
 \end{equation}
 is the Arnowitt–Deser–Misner (ADM) mass, where $h$ denotes the event-horizon radius.
 The classical Schwarzschild solution~\cite{Schwarzschild:1916,Eddington:1924pmh,Finkelstein:1958zz,Kruskal:1959vx,Szekeres:1960gm,Dafermos:2021cbw} is recovered from~\eqref{mtransform} by taking $m(r)={\cal M}$ for all $0<r<\infty$.
 
 To ensure smooth continuity of the metric~\eqref{metric} across $r=h$, the mass function must satisfy
 \begin{equation}
 	\label{cond2}
 	m(h)={\cal M}\ ; \qquad m'(h)=0\ ,
 \end{equation}
 where $F(h)\equiv\,F(r)\big\rvert_{r=h}$ for any $F(r)$. Condition~\eqref{cond2} guarantees that the metric is ${\cal C}^{1}$ across the horizon. 
 
 To construct an interior solution beyond the trivial case $m(r)={\cal M}$, one must adopt a generic form for $m(r)$ subject to the continuity conditions in~\eqref{cond2}. If, in addition, we demand that the interior reduces to Minkowski spacetime ($m\rightarrow\,0\,\text{as}\,\,r\rightarrow\,0$), and exhibits some form of regularity at the origin $r=0$ as measured by curvature scalars, then it becomes difficult to avoid the trivial solution $m(r)={\cal M}$, unless (i) the interior region is extended beyond the Kerr–Schild class~\cite{Mars:1996khm}, or (ii) additional parameters beyond the total mass ${\cal M}$ are introduced~\cite{Maeda:2024tpl}. 
 
 Nevertheless, as we show below, it is possible to construct non-trivial solutions while remaining within the Kerr–Schild class and without introducing extra charges in the interior configuration.
 Although this may appear counterintuitive, a more general interior can indeed be obtained by imposing higher differentiability and requiring the mass function to belong to the class ${\cal C}^{N}$~\cite{Ovalle:2024wtv,Casadio:2024fol,Aoki:2024dyr,Casadio:2025pun}. A convenient starting point is a generic solution expressed as a superposition of $N$ different contributions~\cite{Ovalle:2017fgl,Ovalle:2019qyi} embedded in a de Sitter background with cosmological constant $\sim\,C_3$
 \begin{equation}
	\label{mpoly}
	m(r)
	=
	C_3\,r^3
	+
	\underbrace{C_l\,r^l
		+C_n\,r^n
		+C_p\,r^p+...}_{\text{$N$ terms}}
	\ ,
\end{equation}
 The coefficients $C{s}$ are fixed by~\eqref{cond1} together with
 \begin{equation}
 	\label{cond-n}
 	\frac{d^{n} m}{dr^{n}}(h)=0\ ,
 \end{equation}
 for every $1\le n\le N$. Under these conditions, the interior Schwarzschild mass function becomes
 \begin{eqnarray}
	\label{minfi}
	&&	m(r)=\frac{r}{2}\left[\left(\frac{r}{h}\right)^2\,\prod_{i=1}^{N}\frac{n_i+1}{n_i-2}\right.\nonumber\\
	&&\left.+3(-1)^N\,\sum_{k=1}^{N}\frac{1}{n_k-2}\left(\frac{r}{h}\right)^{n_k}\prod_{\substack{i=1\\i\neq k}}^{N}\frac{n_i+1}{n_k-n_i}\right]\ ,
\end{eqnarray} 
 which yields the corresponding interior metric function~\cite{Ovalle:2025pue}
\begin{eqnarray}
	\label{finfi}
	&&	f(r)=1-\left[\left(\frac{r}{h}\right)^2\,\prod_{i=1}^{N}\frac{n_i+1}{n_i-2}\right.\nonumber\\
	&&\left.+3(-1)^N\,\sum_{k=1}^{N}\frac{1}{n_k-2}\left(\frac{r}{h}\right)^{n_k}\prod_{\substack{i=1\\i\neq k}}^{N}\frac{n_i+1}{n_k-n_i}\right]\ ,
\end{eqnarray} 
 where $2<n_i\in\mathbb{N}$. For each fixed $N$, the solution is characterized by the set $n_i=\{n_1, n_2,\ .\ .\ .n_N\}$, which parametrizes an infinite family of regular Schwarzschild BHs. This family depends solely on the total mass ${\cal M}$ of the configuration and therefore carries no primary hairs. 
 
As a concrete illustration of the mass function in Eq.~\eqref{minfi}, let us consider the simplest case with $N=1$, \footnote{In this case the term $i\neq\,k$ produces an empty product and therefore evaluates to $1$.} 
 which takes the explicit form
 \begin{equation}
 	\label{m1x}
 	m(r)=\frac{r}{2(n-2)}\left[\frac{r^2}{h^2}\left(n+1\right)-3\left(\frac{r}{h}\right)^n\right]\ ;\,\,\,n>2\ ,
 \end{equation}
 and the case $N=2$
 \begin{eqnarray}
 	\label{Mr}
 	m(r)=&&\frac{r}{2}\left[\frac{(n+1)(l+1)}{(n-2)(l-2)}\left(\frac{r}{h}\right)^2+\frac{3\,(l+1)}{(n-2)(n-l)}\left(\frac{r}{h}\right)^n\right.
 	\nonumber\\
 	&&\left.+\frac{3\,(n+1)}{(l-2)(l-n)}\left(\frac{r}{h}\right)^l \right]\ ;\,\,\,l>n>2\,\in\mathbb{N}\ .
 \end{eqnarray}

 We emphasize that the expression for $m(r)$ in Eq.~\eqref{minfi} should be viewed as a minimal ${\cal C}^{N}$ construction satisfying regularity at the origin, $m'(r)>0$ as required for the null convergence condition, and smooth matching at $r=h$. In the limit $N\to\infty$ it extends to a series representation of smooth mass functions, thereby capturing a broad range of physically reasonable collapse profiles.

 Notice that from Eq.~\eqref{finfi} we can identify the effective cosmological constant as
 \begin{equation}
 	\label{Lambda-effec}
 	\Lambda_{eff}=\frac{3}{h^2}\prod_{i=1}^{N}\frac{n_i+1}{n_i-2}
 \end{equation}
which shows that the regular Schwarzschild BH approaches a de Sitter interior when
 \begin{equation}
 	\label{extremal}
 	m(r) \to \frac{r^3}{2h^2}, \quad
 	f(r) \to 1 - \frac{r^2}{h^2}, \quad h_{\rm c} \to h \quad \text{as} \quad n_i \to \infty\ ,
 \end{equation}
 so that the inner horizon $h_{\rm c}$ merges with the event horizon, $h_{\rm c}\sim\,h$, yielding a quasi extremal configuration.

\subsection*{Dynamical evolution}

To analyze the dynamical evolution of the inner Schwarzschild geometries described in Eq.~\eqref{finfi}, we first extend these solutions to time-dependent configurations. A convenient starting point is to rewrite the static metric~\eqref{metric} in Eddington–Finkelstein form by introducing the ingoing null coordinate $v$, defined through $dv=dt+dr/f$, which leads to
\begin{equation}
	\label{EF-metric}
	ds^{2}
	=
	-f(v,r)\,dv^{2}+2\,dv\,dr+r^2\,d\Omega^2\ ,
\end{equation}
where
\begin{equation}
	\label{mtransform2}
	f(v,r)=\left\{
	\begin{array}{l}
		1-\frac{2\,m(v,r)}{r}
		\ ,
		\quad
		{\rm for}\
		0< r \leq h
		\\
		\\
		1-\frac{2\,{\cal M}}{r}
		\ ,
		\quad
		{\rm for}\
		r>h
		\ .
	\end{array}
	\right.
\end{equation}
The time-dependent mass function $m(v,r)$ in~\eqref{mtransform2} is then given by the same expression as~\eqref{minfi}, where the dynamical evolution is mediated exclusively through variations of the parameters $n_i$
\begin{equation}
	\label{ni(t)}
	n_i=n_i(v)\ ,
\end{equation}
which are not required to be integer valued. Since the total mass ${\cal M}$ is conserved, the horizon radius $h=2{\cal M}$ remains time-independent, that is, $h\neq\,h(v)$. Consequently, the mass function in Eq.~\eqref{minfi} now reads
\begin{eqnarray}
	\label{minfi-v}
	&&	m(v,r)=\frac{r}{2}\left[\left(\frac{r}{h}\right)^2\,\prod_{i=1}^{N}\frac{n_i(v)+1}{n_i(v)-2}+3(-1)^N\times\right.\nonumber\\
	&&\left.\,\sum_{k=1}^{N}\frac{1}{n_k(v)-2}\left(\frac{r}{h}\right)^{n_k(v)}\prod_{\substack{i=1\\i\neq k}}^{N}\frac{n_i(v)+1}{n_k(v)-n_i(v)}\right]\ ,
\end{eqnarray} 
which continues to satisfy conditions~\eqref{cond2} and~\eqref{cond-n}

We note that, in order to keep our treatment as general as possible, we impose no particular physical theory and therefore do not make use of equations of motion. Within this framework, the term ``evolution'' refers to spacetime itself, namely, the kinematics of its geometry and of its essential characteristics, such as the inner horizon $h_{\rm c}=h_{\rm c}(v)$. The line-element~\eqref{EF-metric} is well-known in the context of general relativity~\cite{Husain:1995bf,Wang:1998qx}. In this setting, for instance, the mass function~\eqref{minfi-v} gives rise to a configuration characterized by a non uniform energy density $\rho=\rho(v,r)$, anisotropy $\Delta\equiv\,p_\theta-p_r\neq\,0$ with non zero radial pressure $p_r(v,r)$ and tangential pressures $p_\theta(v,r)$, and a non vanishing energy flux $\epsilon=\epsilon(v,r)$.

In contrast, when the metric is considered purely from a geometric standpoint, without assuming any specific dynamics~\cite{Borissova:2025msp,Ovalle:2025pue,Borissova:2025hmj}, the null convergence condition
\begin{equation}
	\label{NCC}
R_{\mu\nu}\l^\mu\l^\nu\geq\,0\ ,\quad\l^{\mu}\l_{\mu}=0
\end{equation}
implies
\begin{equation}
	\label{NCC2}
	\dot{m}\geq\,0
\end{equation}
where $\dot{m}\equiv\frac{\partial\,m}{\partial\,v}$. 

A straightforward analysis of Eq.~\eqref{minfi-v} reveals that the interior geometry becomes singular whenever $n_k(v)=n_i(v)$ for any pair of functions. To avoid such singularities, the evolution of all $n_i(v)$ must evolve without intersections. This requires an initial ordering at $v=v_0$, namely,
\begin{equation}
	\label{order}
	n_1(v_0)<n_2(v_0)<. . . <n_N(v_0)\ ,
\end{equation}
which must be preserved for all $v$ through the constraint
\begin{equation}
	\label{order2}
	\dot{n}_i(v)\leq\dot{n}_j(v)\ ,\quad\forall\,\, i<j\ .
\end{equation}
[The reversed ordering similarly requires $\dot{n}_i(v)\geq\dot{n}_j(v)$.] 

Finally, we can summarize the main features of the dynamical evolution of the Schwarzschild BH interior as follows~\cite{Ovalle:2025pue}:
\begin{itemize}
	\item For $N>1$, the evolution is highly constrained, since synchronized parameter evolution is a necessary (though not sufficient) condition to prevent the appearance of singularities.
	
		\item Two physically distinct dynamical regimes can be identified: negative slopes ($\dot{n}_i < 0$) correspond to collapse, whereas positive slopes ($\dot{n}_i > 0$) describe expansion.
		
		\item Expansion-dominated regimes ($\dot{n}_i > 0$) necessarily violate the null convergence condition~\eqref{NCC2}, while in collapse-dominated regimes ($\dot{n}_i < 0$) this condition is trivially satisfied. 
	
	\item Although the inner horizon $h_{\rm c}$ becomes a dynamical hypersurface, $h_{\rm c} = h_{\rm c}(v)$, it still acts as a Cauchy horizon. 
\end{itemize}

\section{Singularity formation and Minkowski breaking} 
\label{sec3}
\noindent We now examine the limits of the configuration~\eqref{finfi} that arise for specific choices of the parameter set $\{n_i\}$. A straightforward analysis shows that the geometry describes a regular black hole (RBH) whenever
\begin{equation}
	\label{RBH}
	\forall\quad i\quad n_i>2 \quad \iff\quad \text{RBH}\ .
\end{equation}
Because our goal is to explore singularity formation, it is necessary to extend the parameter domain to include values with $n_i\leq 2$. In particular, we will focus on the range $n_i \in [-2,2]$. Within this interval, the spacetime describe BHs with integrable singularities (IS)~\cite{Lukash:2013ts,Ovalle:2023vvu,Arrechea:2025fkk}, that is, configurations in which the curvature scalar $R$ diverges at most as
\begin{equation}
	\label{Rint}
	R\sim\,r^{-2}\ .
\end{equation}
Hence,
\begin{equation}
	\label{IS}
	\exists\quad i\text{ such that }n_i\in [-2,2]\iff\text{IS}\ .
\end{equation}
Within this domain of integrable singularities, if there exists an index $i$ such that $n_i=-1$, the standard singular Schwarzschild solution (SCH) is recovered,
\begin{equation}
	\label{SCH}
	\exists\quad i\text{ such that }n_i=-1 \iff\text{SCH}\ ,
\end{equation}
for which the effective cosmological constant in~\eqref{Lambda-effec} vanishes. This can be seen explicitly by taking, for instance, $n=-1$ in Eqs.~\eqref{m1x} and~\eqref{Mr}. 

If instead some $n_i<-2$, the spacetime develops non-integrable or strong singularities (SS),
\begin{equation}
	\label{SS}
	\exists\quad i\text{ such that }n_i<-2 \iff\text{SS}\ .
\end{equation}
\begin{table*}
	\caption{Behavior of the Kretschmann scalar $\mathcal{K}$ for various values of the parameter $n$, corresponding to integrable singularities, in the case $N=1$ of Eq.~\eqref{finfi}.
		\label{tab}}
	\begin{ruledtabular}
		\begin{tabular}{ c c c c c}
			$n$ & Scaling &Notes& $\Lambda$ &Type of Singularity
			\\
			\hline\hline
			$2$
			&
			$\mathcal{K} \sim [\log(r/h)]^2/h^4$  &Locally Minkowski  for $r\sim\,0$ & $\text{AdS}$ &Timelike
			\\
			\hline
			$1$
			&
			$\mathcal{K} \sim (hr)^{-2}$   &Locally Minkowski  for $r\sim\,0$ &$\text{AdS}$ &Timelike
			\\
			\hline
			$0$
			&
			$\mathcal{K} \sim r^{-4}$  &MB $\Rightarrow$ Cauchy horizon $h_{\mathrm{c}}$ vanishes &$\text{AdS}$  &Spacelike
			\\ 
			\hline
			$-1$ & 
			$\mathcal{K} \sim h^2 / r^6$
			&SCH $\Rightarrow$ Point-source solution & $0$ &Spacelike
			\\ 
			\hline
			$-2$ & 
			$\mathcal{K} \sim h^4 / r^8$ & dS+RN-like  ($Q^2<0)$ $\Rightarrow$ dS+Conformal solution & $\text{dS}$ &Spacelike
			\\  
		\end{tabular}
	\end{ruledtabular}
\end{table*}
\begin{figure}
	\includegraphics[width=0.45\textwidth]{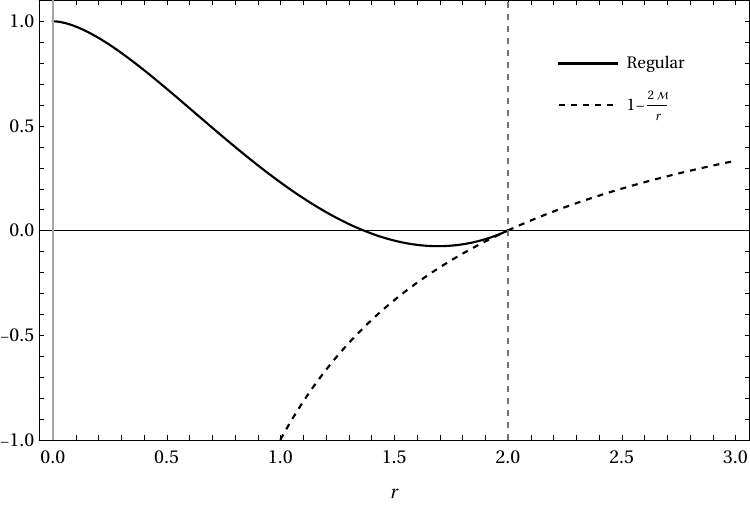}  \
	\caption{\footnotesize Minkowski breaking for the lapse function $f(v,r)$. Since the event horizon must form before the singularity evolves, in agreement with the weak Cosmic Censorship Conjecture, the formation of a regular configuration with its corresponding inner horizon is unavoidable prior to the emergence of the singularity, as represented by the solid line. The dashed curve corresponds to the singular Schwarzschild solution. In order to transition continuously from the regular configuration to the Schwarzschild solution, the condition $f(v,0)=1$ must be abandoned. At present, however, we are not aware of any dynamical mechanisms that could explain this transition. The radial coordinate $r$ is expressed in units of ${\cal M}$.}
	\label{fig3}
\end{figure}
Finally, and this point will be important in what follows, within the integrable domain of~\eqref{IS}, if there exists an index $i$ such that $n_i=0$, the Minkowski metric is not recovered as $r\rightarrow\,0$, in other words, the metric is not locally $\eta_{ab}=\text{diag}(-1,1,1,1)$. This ``Minkowski breaking'' (MB) can be expressed as 
\begin{equation}
	\label{MB}
	\exists\quad i\text{ such that }n_i=0 \iff\text{MB}\ .
\end{equation}
Thus, within the integrable singularity domain, there exist a subrange $n_i \in (0,2]$ where a locally Minkowskian frame can still be defined, whereas for $n_i \in [-2,0]$ the singularity, though integrable, prevents the existence of a smooth local Lorentz frame. See Figure~\ref{fig3} for a general, model independent discussion, and Table~\ref{tab} for specific details of the present construction.

\begin{figure}
	\includegraphics[width=0.47\textwidth]{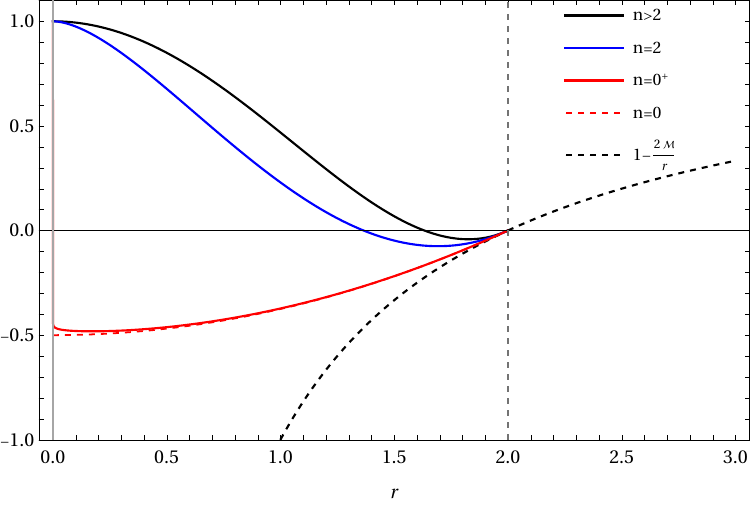}  \
	\caption{\footnotesize Evolution of the metric function~\eqref{sol1} from a regular BH configuration with $n(v)>2$ down to the critical threshold $n(v)=2$. The evolution then proceeds to a second threshold, $n(v)=0$, marking the onset of Minkowski breaking $[f(v,0)\neq1]$, and thereby signaling the loss of spacetime continuity. The horizon remains fixed at $h=2{\cal M}\neq h(v)$. The radial coordinate $r$ is expressed in units of ${\cal M}$.}
	\label{fig1}
\end{figure}

Having established the static structure and the limits associated with integrable and non-integrable singularities, we now turn to the dynamical regime to understand how these singular behaviors emerge during collapse. To this end, we focus on the first term in the ordering~\eqref{order}, namely $n_1(v)\equiv n(v)$, which plays the dominant role. As a consequence of the constraint~\eqref{order2}, this term remains the smallest throughout the evolution and therefore is the one responsible for triggering the onset of singular behavior, being consistently the closest to the critical value $n=2$. To simplify the analysis without loss of generality, we restrict our attention to the case $N=1$,
for which the dynamical version of~\eqref{m1x} takes the explicit form
\begin{equation}
	\label{m1}
	\hspace*{-1.1mm}	m(v,r)=\frac{r}{2[n(v)-2]}\left[\frac{r^2}{h^2}\left[n(v)+1\right]-3\left(\frac{r}{h}\right)^{n(v)}\right],
\end{equation}
which yields
\begin{equation}
	\label{sol1}
	\hspace*{-3.3mm}	f(v,r)=1-\frac{1}{[n(v)-2]}\hspace*{-1.3mm}\left[\frac{r^2}{h^2}\left[n(v)+1\right]-3\left(\frac{r}{h}\right)^{n(v)}\right].
\end{equation}
We can now describe in detail the evolution of the regular Schwarzschild BH. Since the null convergence condition~\eqref{NCC} enforces $\dot{n}_i < 0$ and therefore collapse, the eventual onset of singularities is unavoidable.\footnote{Unless an \textit{ad hoc} regularization mechanism is introduced~\cite{Ovalle:2025pue}.} This behavior is clearly illustrated in Figure~\ref{fig1}, which shows the evolution beginning from an initially regular configuration with $n > 2$, passing through the singular threshold $n=2$, and eventually reaching the Schwarzschild limit $n=-1$. As the figure makes evident, during this dynamic process an unavoidable discontinuity arises at $r=0$, characterized by 
\begin{equation}
\label{jump}
f(v,0)=\left\{
\begin{array}{l}
	1
	\ ,
	\quad
	{\rm for}\ n=0+\delta\ ,\,\,\,\,\,\delta\ll\,1 
	\\
	\\
	-0.5
	\ ,
	\quad
	{\rm for}\	n=0 .
\end{array}
\right.
\end{equation}
This allows us to conclude that the ``Minkowski breaking'' identified in the static case through Eq.~\eqref{MB} corresponds to a state that cannot be realized continuously. On the contrary, it manifests as a discrete transition.

It is instructive to examine in more detail what happens as $n \rightarrow 0$. Let us begin by writing the location of the local extremum (minimum) of the metric function~\eqref{sol1}
\begin{equation}
	\label{extrema2}
	r_{\mathrm{e}}(v)=h\left[\frac{2}{3}\left(1+\frac{1}{n(v)}\right)\right]^{\frac{1}{n(v)-2}}\ .
\end{equation}
As $n(v)$ approaches zero, this extremum moves steadily toward the origin $r=0$, causing the Cauchy horizon $h_{\mathrm{c}}$ to shrink correspondingly,
\begin{equation}
	\label{extremal2}
	r_{\mathrm{e}}(v) \to 0, \quad
	h_{\mathrm{c}}(v) \to 0 \quad \text{as} \quad n(v) \to 0\ .
\end{equation}
Indeed, the dynamical Cauchy horizon $h_{\mathrm{c}}(v)$ vanishes precisely at $n=0$, as shown in Figure~\ref{fig1}. We therefore conclude that the collapse process can be described continuously only up to the disappearance of the Cauchy horizon, beyond which the very continuity of spacetime itself ceases to exist.

\begin{figure}
	\includegraphics[width=0.45\textwidth]{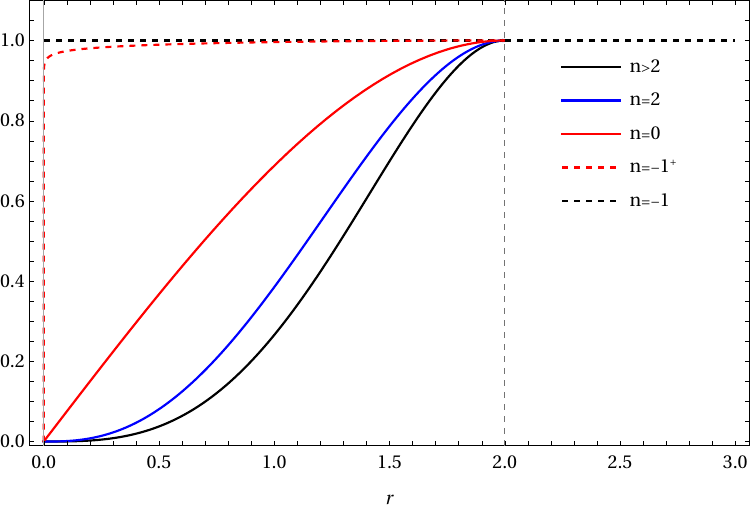}  \
	\caption{\footnotesize Evolution of the mass function~\eqref{m1} from a regular BH configuration with $n(v)>2$ down to the critical threshold $n(v)=2$. The evolution then continues until a second threshold is reached at $n(v)=-1$, where a sudden collapse of all matter ${\cal M}$ into $r=0$ occurs, producing the point-like solution (Schwarzschild). The horizon remains fixed at $h=2{\cal M}\neq h(v)$. The radial coordinate $r$ is expressed in units of ${\cal M}$.}
	\label{fig2}
\end{figure}
If we now examine the total mass ${\cal M}$, which is initially distributed throughout the entire interior region, we find that it also exhibits a discontinuous behavior at $r=0$. However, this discontinuity does not occur during the Minkowski breaking described in~\eqref{jump}, but rather after it, namely,
\begin{equation}
	\label{jump2}
	m(v,0)=\left\{
	\begin{array}{l}
		0
		\ ,
		\quad
		{\rm for}\ n=-1+\delta\ ,\,\,\,\,\,\delta\ll\,1 
		\\
		\\
		{\cal M}
		\ ,
		\quad
		{\rm for}\	n=-1 .
	\end{array}
	\right.
\end{equation}
This shows that the collapse of the total mass ${\cal M}$ into the region $r=0$, that is, the formation of the point-like solution (Schwarzschild), does not occur continuously either. In fact, there is never a gradual accumulation of mass at $r=0$ or any progressive point-source formation, but rather a sudden collapse of all matter ${\cal M}$ into that point, as shown in Figure~\ref{fig2}. This discontinuous process occurs at $n(v)=-1$, that is, after the breakdown of spacetime continuity at $n(v)=0$.

We conclude by emphasizing that our model is simple in one respect, as it assumes the existence of the event horizon from the outset, yet highly nontrivial in another, since it allows for the construction of an extremely rich variety of internal geometries consistent with this assumption. It is precisely this richness that makes it possible to describe, in a fully analytical manner, several features of the final stages of collapse. Remarkably, all of this is achieved within a single subclass of spacetime, namely the Kerr–Schild type, valid throughout the entire spacetime region $0\leq\,r\,\leq\infty$. If, in order to compare with previous studies, one specializes to general relativity, it becomes clear that the nonuniform and anisotropic collapsing fluid present in our model lies well outside the restrictive settings usually considered, such as homogeneous dust collapse~\cite{Oppenheimer:1939ue,Shibata:1999va}, inhomogeneous dust collapse~\cite{Tolman:1934za,Bondi:1947fta,Lemaitre:1933gd,Christodoulou:1984mz,Joshi:1993zg,Joshi:2001xi}, isotropic~\cite{Lasky:2006mg,Mosani:2020ena} or  anisotropic configurations~\cite{Mena:2004ck} that do not belong to the Kerr–Schild class.

\section{Final remarks.} 
\noindent Gravitational collapse, like all classical processes, unfolds as a continuous evolution. Consequently, the geometry of spacetime accompanying the collapsing matter should also evolve smoothly. The classical description of BH formation ought to follow this same principle. Yet understanding where and how this continuity fails is crucial, as it marks the boundary of classical physics, the point beyond which spacetime itself demands a new, fundamentally different description.

A direct comparison between the lapse functions $f(v,r)$ for regular and Schwarzschild BHs shows that, to pass continuously from one to the other, the condition $f(v,0)=1$ [or equivalently $m(v,0)=0$] must be abandoned (see Figure~\ref{fig3}). This signals the onset of what we have termed ``Minkowski breaking'', the moment when the smooth fabric of spacetime ceases to exist. Constructing a dynamical model that illustrates this transition is certainly important, but our results suggest something more profound, namely, that such a transition may not occur conti\-nuously.

In essence, the Schwarzschild BH could emerge from a regular configuration only through a discrete change in the fundamental structure of spacetime. A general proof of this behavior, which  would require constructing an interior solution even more general than~\eqref{finfi}, if such a construction is possible, would point unmistakably to a deeper principle, namely the discretization of spacetime as the natural and necessary framework for describing both the formation and the regularization of gravitational singularities.

\subsection*{Acknowledgments}
\vspace*{1mm}
JO thanks the Centro de Estudios Cient\'ificos (CECs) for their support throughout the development of this research. This work was partially supported by ANID–FONDECYT Grant No. 1250227.
%

%
%
%
\bibliography{references.bib}
\bibliographystyle{apsrev4-1.bst}
%
%
\end{document}